\documentclass[10pt,twocolumn,superscriptaddress,floafix,nobalancelastpage]{revtex4-1}
\usepackage{amsmath,graphicx,times,xfrac,relsize,float,mathtools,amssymb}

\usepackage[unicode=true,breaklinks=false,pdfborder={0 0 1},backref=false,colorlinks=true,linkcolor=blue,citecolor=blue]{hyperref}

\AtBeginDocument{\usepackage{booktabs}}

\newcommand{\ds}{\displaystyle}

\usepackage[normalem]{ulem}

\newcommand{\bra}[1]{\left\langle{#1}\right|}
\newcommand{\ket}[1]{\left|{#1}\right\rangle}
\newcommand{\ip}[2]{\left\langle{#1}\middle|{#2}\right\rangle}                  
\newcommand{\bk}[3]{\left\langle{#1}\middle|{#2}\middle|{#3}\right\rangle}      
\newcommand{\exx}[1]{\left\langle{#1}\right\rangle}                             

\renewcommand{\epsilon}{\varepsilon}
\newcommand{\w}{\omega}
\newcommand{\e}{\epsilon}

\newcommand{\rmi}{\text{i}}

\usepackage{dsfont}
\newcommand{\id}{\mathds{1}} 

\newcommand{\ssub}[1]{{\protect\scalebox{0.9}{$_#1$}}}
\newcommand{\R}{\ssub{R}}
\newcommand{\I}{\ssub{I}}
\newcommand{\Gs}{\ssub{G}}
\newcommand{\hint}{H^\text{int}}
\newcommand{\hintp}{{H'}^{\text{int}}}
\newcommand{\B}{B}

\newcommand{\kp}[2]{\psi^{\scalebox{0.75}{(#1)}}_{#2}}
\newcommand{\kc}[2]{\chi^{\scalebox{0.75}{(#1)}}_{#2}}

\renewcommand{\O}{\mathcal{O}}
\newcommand{\PSI}{\ket{\scalebox{1.2}{${\Psi}$}_{0,-}}}

\begin{document}

\title{A Simple Test of the Equivalence Principle(s) for Quantum Superpositions}
\author{Patrick J. Orlando}
\affiliation{School Of Physics \& Astronomy, Monash University, Clayton, Australia 3800}
\author{Robert B. Mann}
\affiliation{Department of Physics and Astronomy, University of Waterloo, 200 University Avenue West, Waterloo, Ontario N2L 3G1, Canada}
\affiliation{Perimeter Institute for Theoretical Physics,
31 Caroline Street North, Waterloo, Ontario N2L 2Y5, Canada}
\author{Kavan Modi}
\affiliation{School Of Physics \& Astronomy, Monash University, Clayton, Australia 3800}
\author{Felix A. Pollock}
\email{felix.pollock@monash.edu}
\affiliation{School Of Physics \& Astronomy, Monash University, Clayton, Australia 3800}

\begin{abstract}
We propose a simple experimental test of the quantum equivalence principle introduced by Zych and Brukner [arXiv:1502.00971], which generalises the Einstein equivalence principle to superpositions of internal energy states. We consider a harmonically-trapped spin-$\frac12$ atom in the presence of both gravity and an external magnetic field and show that when the external magnetic field is suddenly switched off, various violations of the equivalence principle would manifest as otherwise forbidden transitions. Performing such an experiment would put bounds on the various phenomenological violating parameters. We further demonstrate that the classical weak equivalence principle can be tested by suddenly putting the apparatus into free fall, effectively `switching off' gravity.
\end{abstract}

\maketitle

{\bf Classical equivalence principles.---}
At least since the days of Galileo and Newton, it has been known that acceleration under gravity is independent of an object's mass~\cite{Galilei:1953ua, Newton:821668}. This peculiarity has led to the proposition of gravitational equivalence principles which, if broken, represent a departure from our current understanding of gravity. The weak equivalence principle ({\bf WEP}) states that all objects, starting with the same position and velocity, and subject only to a gravitational field will follow the same trajectory, irrespective of the object's constituents or properties~\cite{WILL:1984ef}. Mathematically, this statement translates to the inertial prescription of mass being equal to the gravitational prescription. Early tests of the WEP consisted of dropping similar objects of differing mass or measuring the period of pendulums~\cite{Roll:1964cm}. These have now been superseded by extremely accurate torsion balance experiments \cite{Roll:1964cm, VonEotvos:1890tn, Wagner:2012vh}, whose null results place limits as stringent as one part in $10^{13}$ on WEP violation. Hoping to detect violations at very low mass scales, free-fall experiments have been performed with systems as light as individual neutrons, finding no deviation from equivalence~\cite{McReynolds:1951dv, Dabbs:1965bo, Koester:1976et, Sears:1982il}. 

The mass-energy relation of special relativity~\cite{Einstein:1911wk},  $E=mc^2$, dictates that internal energy of a system must contribute to its mass; different internal energy states correspond to different \emph{effective masses} for the system: $m_k = m_k^\text{ext} + E_k^\text{int}/c^2$, where the index $k = \{R,I,G\}$ denotes quantities corresponding to the rest, inertial, and gravitational masses respectively and $m^\text{ext}$ is the mass of the system when the internal energy is at its lowest. The `rest' internal energy $E_R^\text{int}$ is derived from the Hamiltonian that generates internal dynamics in the absence of external motion. By explicitly labelling the inertial and gravitational contributions, we examine the possibility that they differ from each other. The gravitational effects of the mass-energy equivalence can be measured in the weak-field limit---where a Newtonian description is mostly satisfactory---avoiding the need for the complete machinery of general relativity. An example is the gravitational redshift observed in the seminal Pound-Rebka experiment, which can be calculated by coupling the effective mass $\tilde{m} = h\nu/c^2$ to a Newtonian potential~\cite{Pound:1960iz}.

The condition $m_\R = m_\I = m_\Gs$ embodies the Einstein equivalence principle ({\bf EEP}), which requires the mass-energy of a system to be a universal property for all observers. The EEP can be deconstructed into three distinct conditions: (a) WEP holds, i.e., $m_\I = m_\Gs$, which implies that the internal energy contributes equally to inertia and weight; (b) the validity of special relativity, or local Lorentz invariance ({\bf LLI}), which requires internal energy to contribute equally to both inertia and rest mass; (c) measurement outcomes are independent of their position in spacetime, or local position invariance ({\bf LPI}); a consequence of LPI is that internal energy contributes equally to both weight and rest mass \cite{WillCMBook}. If any two of WEP, LLI, and LPI are satisfied, than the third is also satisfied. However, there is a yet stronger, quantum equivalence principle, proposed by Zych and Brukner \cite{Zych:2015vm}, that may not hold even if the three principles above do.

{\bf Quantum equivalence principle.---}
One of the major goals of modern physics is to identify gravitational effects in quantum mechanics that are not present in classical theory. This has motivated several equivalence principle tests using quantum systems\cite{Bonnin2015, Williams2015, Zhou2015, Hartwig2015}, where the effect of gravity on spatial superpositions can be probed. In each of these experiments, the quantum system in question is a cold gas of two different atomic species, whose relative motion is used to test for violations. 

Here, on the other hand, rather than using atoms of different masses, we propose a test which distinguishes between the gravitational motion of different internal energy states of the \emph{same} particle. This introduces an extra richness to the problem, since, due to the possibility of superposition, the particle's internal energy does not always have a well defined value. Following Ref.~\cite{Zych:2015vm}, we treat this by promoting the mass to an operator of the form,
\begin{gather}
M_k = m_k \openone^\text{int} + \frac{1}{c^2}H_k^\text{int},\label{eq:MassOp}
\end{gather}
where $k = \{R,I,G\}$ again refers to rest, inertial and gravitational masses; $m_k$ is the mass of the system when the internal energy is its lowest energy eigenstate; and the internal Hamiltonians encapsulate the effective contributions to the respective masses. Once more, the subscripts allow for the possibility that the effective weight and inertia of the internal energy may differ, e.g., $H_\I^\text{int}$ and $H_\Gs^\text{int}$ are not \textit{a priori} equivalent. Classical tests of the equivalence principle, and those described earlier in this section, only probe systems in eigenstates of their internal Hamiltonian, thereby testing the diagonal elements of the mass operator in Eq~\eqref{eq:MassOp}. Whereas, the quantum version of EEP is expressed in terms of operator equations: namely, LLI requires $M_\I =M_\R$, LPI requires $M_\Gs =M_\R$ and WEP requires $M_\I =M_\Gs$. These impose additional constraints on the off-diagonal elements of the operators, which allow for the equivalence principle to hold classically whilst being violated for superpositions of internal energy.

For a spin-$\frac{1}{2}$ atom, the internal Hamiltonian in the presence of an external field of magnitude $\B$ is,
\begin{gather}
\hint_\R = \mu \B \ket{\textstyle{+\frac12}}\bra{\textstyle{+\frac12}}. \label{H-int-r}
\end{gather}
To avoid negative contributions to the mass-energy of the system, the energy splitting is defined asymmetrically such that the lowest eigenvalue is zero (we elaborate on this point in the concluding remarks). In general, we do not assume that the internal energy contributes identically to rest, inertial and gravitational mass. We encompass this by introducing equivalence violating operators $\xi_\I$ and $\xi_\Gs$, which respectively quantify deviations from LLI and LPI per unit energy. These $\xi_k$ then satisfy,
\begin{gather}
H^\text{int}_k = \hint_\R + \mu \B\;\xi_k, \quad
\xi_k = \left[\begin{array}{cc}a_k & b_k\\ b^*_k & c_k \end{array}\right].\label{eq:Deviations}
\end{gather}
When defining the violating parameters $\xi_k$, we have assumed a linear scaling with internal energy. Any non-linear violation would manifest as different values of the $\{a_k,b_k,c_k\}$ for different external field strengths.

In this Letter, we propose an experiment to probe the quantum equivalence principle using a harmonic oscillator with two internal levels. Using the mass operator formalism, we calculate transitions caused by the induced coupling between internal and external degrees of freedom and find that certain transitions are only allowed when either WEP, LLI, or LPI is violated. We go on to discuss the sensitivity of a candidate system to these violations and the relative strength of thermal noise.

{\bf Setup.---}
In order to design the simplest possible test of the quantum equivalence principle, we consider the situation where a spin-$\frac{1}{2}$ atom is trapped in a harmonic potential.  This has the advantage of being both mathematically tractable and physically realisable. Most physical systems behave like a harmonic oscillator close to equilibrium, and (approximate) two-level systems are relatively common. In some cases this may be the ground and excited state of a trapped atom, but in this treatment we require that the energy splitting can be externally controlled and will thus consider a spin-$\frac{1}{2}$ degree of freedom, whose energy we can manipulate with an external magnetic field.

One potential experimental system satisfying these requirements is a spin-$\frac{1}{2}$ atomic isotope held in a magneto-optical trap, as is commonly used in cold atoms experiments \cite{PhysRevX.2.041014}---taking into account the higher spins of many commonly trapped atomic species would be a straightforward generalisation of the results we present here. 
One might worry that, since all atoms have internal structure beyond their ground state spin, that this might also contribute significantly to the effect described in this Letter. However, we show in Appendix~\ref{app:OtherInternal} that any additional internal energy levels do not affect our results, as long as the presence of the external field does not significantly perturb them.

\begin{figure}[t]
\includegraphics[width=0.8\linewidth]{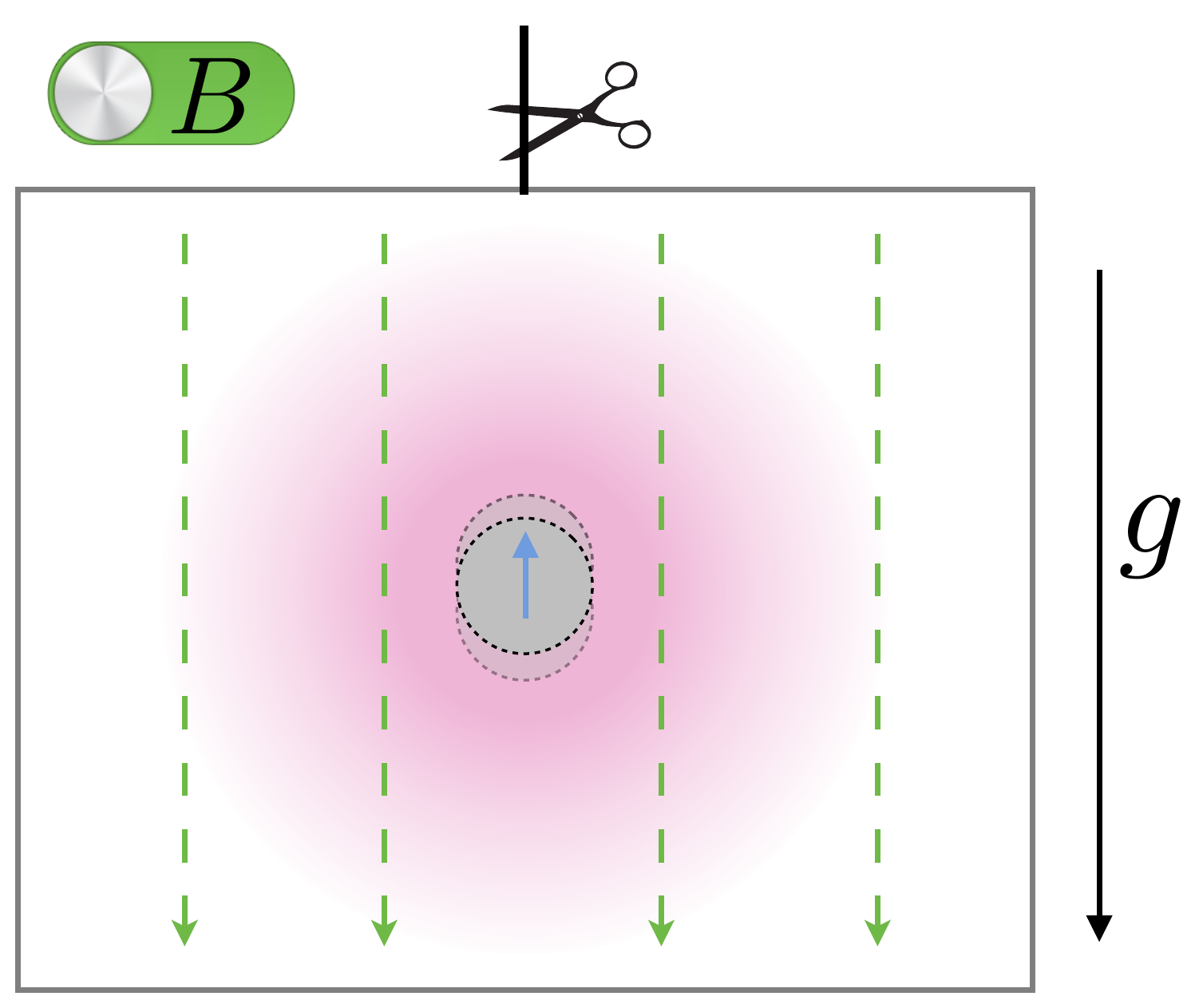}
\caption{\textbf{A diagram of the proposed experiment.} It consists of a trapped atom held in a Gaussian optical trap at some fixed height in the presence of an external magnetic field $\B$. After cooling the spin-$\frac12$ atom to its ground state, the magnetic field is suddenly turned off. In a different experiment the entire enclosure is set into free fall by disconnecting it from the suspension point, effectively transforming away the gravitational potential. In both cases the transition probabilities out of the ground state yield information about the magnitude of violations of different equivalence principles.}
\label{fig:Setup}
\end{figure}

Our proposed tests require the whole experimental apparatus to be suspended at some height $x$ in a gravitational field, and, for some tests, that it may be dropped. This has the effect of ``switching off'' gravity as it freely falls, in accordance with the EEP. A diagram of the setup is shown in Fig.~\ref{fig:Setup}. We begin by writing down the full Hamiltonian, including the mass operators:
\begin{gather}
H= M_\R c^2 + \frac{P^2}{2M_\I } + \frac12 \kappa x^2 + M_\Gs g x.  \label{eq:InitHam}
\end{gather}
The Hamiltonian consists of the rest mass-energy operator $M_\R c^2$, which is no longer completely degenerate; the modified kinetic energy operator, dependent on the inertial mass operator $M_\I$; the harmonic potential energy, with spring constant $\kappa$; and the gravitational potential energy, dependent on gravitational mass operator $M_\Gs$. This consistently associates the appropriate internal Hamiltonian $\hint_k$ with its external coupling, and assumes that the harmonic potential couples to the same degree of freedom as couples to gravity. Henceforth, it will be convenient to define $\w_0 = \sqrt{{\kappa}/{m_I}}$ as the natural frequency of the oscillator when in its lowest internal energy eigenstate. 

After inserting the definition of the mass operator in Eq.~\eqref{eq:MassOp} and approximating $1/M_\I$ through a Taylor expansion, (see Appendix \ref{App:H-Derivation} for details), we arrive at the following approximation for the Hamiltonian:
\begin{align}
H =& D + \underbrace{\hint_\R + \frac{P^2}{2m_\I } + \frac12 m_\I \w_0^2 \left( x+ \nu \frac{g} {\omega_0^2} \right)^2}_{H_0}  \label{eq:FullHam} \\
& + \frac{1}{m_\I c^2}\Bigg\{- \frac{P^2}{2m_\I }\hint_\I+ m_\I gx\hint_\Gs\Bigg\} + \mathcal{O}\left(\frac{1}{m_\I^2 c^4}\right),\nonumber
\end{align}
which is a valid approximation provided that ${\|\hint_\I\|}/{m_\I c^2} \ll 1$. Above, $\nu = m_\Gs/m_\I$ and its deviation from unity indicates a violation of the classical version of the WEP. Finally, terms corresponding to the rest mass energy $m_\R c^2$ and the mean gravitational energy $\frac12 m_\Gs^2g^2/\kappa$ have been collected into the constant $D$. This constant merely generates a global phase, and its value does not affect our results.

We now describe an experiment which can isolate each of the parameters in these equivalence violating operators, thereby testing each of the conditions of the quantum equivalence principle.

{\bf Proposed Equivalence Tests.---}
The additional terms in Eq.~\eqref{eq:FullHam}, introduced by the mass operator, entangle the typically uncoupled spin and oscillator subspaces. In general, the eigenstates of this Hamiltonian do not have a closed form, due to the coupling between internal and external degrees of freedom. However, we use the fact that $m_\I c^2$ is much larger than the internal energy difference $\mu \B$ to define a dimensionless parameter 
\begin{gather}\label{lambda}
\lambda = \frac{\mu \B}{m_\I c^2}.
\end{gather}
The eigenstate of $H$ can be computed as a perturbative expansion in $\lambda$ and the known eigenstates of $H_0$ through the expressions,
\begin{align}
& \ket{\Psi_{n,s}} = \ket{\kp{0}{n,s}} + \lambda \ket{\kp{1}{n,s}}+\mathcal{O}(\lambda^2) , \label{eq:PertExpFormula}\\
&\ket{\kp{1}{n,s}} = \sum_{\{l,r\}\neq \{n,s\}} \frac{\bk{\kp{0}{l,r}} {V}{\kp{0}{n,s}}} {E^{(0)}_{n,s} - E^{(0)}_{l,r}} \ket{\kp{0}{l,r}},
\label{eq:FirstOrderCorrectionFormula}
\end{align}
where $V = (H-H_0)/\lambda$ is the perturbation and the indices $\{n,s\}$ and $\{q,r\}$ enumerate the entire set of unperturbed eigenstates of $H_0$, in terms of their harmonic and spin quantum numbers (see Appendix~\ref{app:FullCalc} for details). The superposition of product states above suggests that if we start in some eigenstate of the full Hamiltonian $\ket{\Psi}$, and measure in the basis of product states $\ket{\kp{0}{n,s}}=\ket{n,s}$ (by effectively decoupling internal and external degrees of freedom), then the probabilities of different outcomes depend on the square of the coefficients in Eq.~\eqref{eq:FirstOrderCorrectionFormula}. The coefficients themselves depend on the perturbation $V$, which has been carefully expressed in Eq.~\eqref{eq:Deviations} to account for the possibility that the internal energy may not couple identically to rest, inertial and gravitational mass. 

In practice, it is difficult to prepare the system in an arbitrary eigenstate of the full Hamiltonian---especially since we are assuming that it comprises unknown parameters---but by cooling the system, we can prepare it in its ground state. The fact that the ground state can be reliably prepared many times makes it an ideal choice for this test. Starting from Eq.~\eqref{eq:FirstOrderCorrectionFormula} we compute the first order corrected ground state:
\begin{align}
\PSI =&  \ket{0,\textstyle{-\frac12}} - \lambda\left[\rule{0cm}{8mm} \left(\frac{\hbar \w_0 b_\I^*}{4\mu \B} + b_\Gs^*\frac{ m_\I \nu g^2}{\mu \B\w_0^2}\right)\ket{0,\textstyle{+\frac12}}\right.\nonumber \\
&+\frac{a_\I }{4\sqrt{2}}\ket{2,\textstyle{-\frac12}}+\frac{b_\I^* }{4\sqrt{2}(1 + \frac{\mu \B}{2\hbar\w_0})}\ket{2,\textstyle{+\frac12}} \label{eq:GroundState}\\
&+ \left. \rule{0cm}{8mm}\sqrt{\frac{g^2 m_\I}{2\hbar\w_0^3}}\left( a_\Gs\ket{1,\textstyle{-\frac12}} + \frac{b_\Gs^*}{1 +\frac{\mu \B}{\hbar\w_0}} \ket{1,\textstyle{+\frac12}}\right) \right].\nonumber
\end{align}
First order corrections to all eigenstates including their calculation can be found in Appendix~\ref{app:FullCalc}. The ground state corrections contain only parameters of the equivalence violating operators $\xi_k$ in Eq.~\eqref{eq:Deviations}. Thus if the system is cooled in the true ground state of $H$ and a measurement is made in the basis of $H_0$, an outcome different from $\ket{0,-\frac12}$ will only be observed if the equivalence principle has been violated in some way (up to first order in $\lambda$). The associated leading-order probabilities are listed in Table~\ref{tab:TransitionProbs}. 

{\bf Measuring the violating amplitudes.---}
The difficulty then lies in being able to measure the state of the oscillator in the basis of the unperturbed states. One method for doing this is as follows: (i) Cool the system into the ground state of the full Hamiltonian $H$. (ii) Turn off the magnetic field and perform an energy measurement on the oscillator to determine its number state. (iii) Perform a spin polarising measurement to determine the spin state. The statistics of these measurements are sufficient to deduce the magnitudes of the $a_k$ and $b_k^*$ terms of the equivalence violating operators $\xi_k$. 

\begin{table}[t]
\begin{ruledtabular}
\begin{tabular}{clcrr}
State &  Probability & Principle & \multicolumn{2}{r}{$\O$(Coefficient)}\\ 
$(n,\;s)$ & $\times (m_\I c^2/\mu \B)^2$ & Tested & $^3$He & $^{171}$Yb
\\ \midrule
 $0,\;+\frac12$ & $\ds\frac14 \left|\frac{\hbar \w_0 }{2\mu \B}b_\I + \frac{ m_\I \nu g^2}{\mu \B\w_0^2}b_\Gs\right|^2$ &  QWEP$^*$ & $10^{-40}$ &$10^{-42}$\\[3mm]
 $1,\; -\frac12$ & $\ds\frac{m_\I  g^2  }{2 \hbar\w_0^3}a_\Gs^2$ & LPI & $10^{-36}$& $10^{-39}$\\[3mm]
 $1,\; +\frac12$ & $\ds\frac{m_\I  g^2  }{2 \hbar\w_0^3}\frac{|b_\Gs|^2}{(1 + \frac{\mu \B}{\hbar \w_0})^2}$ & QLPI & $10^{-44}$& $10^{-46}$ \\[3mm]
 $2,\; -\frac12$ & $\ds\frac{a_\I^2}{32}$& LLI & $10^{-35}$ & $10^{-40}$\\[3mm]
 $2,\; +\frac12$ & $\ds\frac{|b_\I|^2}{32(1+\frac{\mu \B}{\hbar w_0})^2}$ & QLLI & $10^{-43}$& $10^{-46}$\\
\end{tabular}
\end{ruledtabular}
\caption{\textbf{Transition probabilities to different decoupled spin and oscillator states when the external magnetic field is switched off.} Observation of these transitions indicates violation of a related equivalence principle. The parameter $\nu = m_\Gs/m_\I$, if different from unity, represents a classical violation of the WEP, whilst any violation leading to non-zero probabilities comes from considering the relativistic coupling of the internal energy. In addition the violations labelled QLLI and QLPI are unique to quantum mechanical tests, and do not necessarily indicate violation of the corresponding classical principles. If only one of (Q)LLI or (Q)LPI is violated, then a violation of (Q)WEP is implicated. If both violations occur then their size must be calculated in order to confirm that the (Q)WEP is indeed also violated. The probability labelled QWEP$^*$ enables measurement of the relative phase between the two quantum violating parameters. The final columns show order-of-magnitude transition probabilities for atoms in a $1\, {\rm T}$ external field with a typical optical dipole trapping frequency of $\sim$10 kHz; they are per unit violation, i.e., they assume $b_k = a_k = c_k = 1$ $\forall \,k$.}
\label{tab:TransitionProbs}
\end{table}%

We can also determine the relative phase between $b_\I$ and $b_\Gs$ provided we can independently measure the weak equivalence violating parameter $\nu$. To do this, a secondary experiment is required, where the oscillator is prepared in the ground state $\ket{0}$ of the Hamiltonian $H_0$, by cooling in the absence of the external magnetic field---in which case the internal degree is irrelevant. The whole apparatus is then dropped---moving into a freely-falling frame where the value of $g$ is effectively zero. This free fall Hamiltonian $H_f$ is related to $H_0$ by the displacement operator
\begin{gather}
D(\alpha) = \exp[\rmi \sqrt{2}\alpha p]
\quad \mbox{with} \quad
\alpha = \nu g \sqrt{\frac{m_\I}{(2\hbar\w_0^3)}}.
\end{gather}
as
$H_f = D(\alpha)H_0D(-\alpha)$. An energy measurement of the oscillator in the freely falling frame would yield, on average,
\begin{align}
\exx{H_f} =& \bk{0}{D(\alpha)H_0D(-\alpha)}{0}
\nonumber\\
=& \frac{\hbar\w_0}{2} +  m_\I \nu^2 \frac{g^2}{2\w_0^2} , \label{eq:DroppingNoMassOp}
\end{align}
thereby constituting a direct measurement of the weak equivalence violating parameter $\nu$. This allows isolation of the terms $a_k$ and $b_k^*$, but not $b_k$ and $c_k$. The latter parameters only appear in the first order correction to the state $\ket{0,+\frac12}$, which, being the first excited state of $H$, may be difficult to reliably and repeatedly prepare. We have calculated this corrected spin-up state in Appendix~\ref{app:spin}.

Our results are summarised in Table~\ref{tab:TransitionProbs}: the first two columns list the possible transitions and their corresponding probabilities, the middle column indicates which equivalence principle an observed transition violates, and the final two columns list the order of magnitude of different transition probabilities for $^3{\rm He}$ and $^{171}$Yb per unit violation, i.e., we have set $a_\I=a_\Gs=b_\I=b\Gs=1$. These isotopes were chosen since they have a spin-$\frac12$ ground state with sizable magnetic moments of 2.128 and 0.492 respectively. Lastly, we have used reasonable experimental parameters of $\w_0 = 10\, {\rm kHz}$ and $\B= 1\, {\rm T}$ to calculate the order of magnitude estimates. 

{\bf Discussion.---} 
Though we find minuscule transition probabilities \emph{per unit violation}, we cannot \textit{a priori} say anything about the size of the violating parameters. In fact, the suppressing factor of $\lambda^2$ could be obfuscating significant violations at the quantum level, especially when it comes to the parameters $b_\I$ and $b_\Gs$, which have hitherto not been tested experimentally and may be non-zero even if the EEP holds classically. The measurement of any violations would have significant implications for our understanding of quantum gravity. In particular, certain string theories predict a fluctuating violation that scales inversely with mass-energy \cite{Damour1994, Damour2012}, thus preventing its observation in experiments with larger masses.

It can be seen from Table~\ref{tab:TransitionProbs} that, though larger masses have a stronger coupling to gravity, the probability of observing a violation \emph{decreases} with the mass of the atom. This is because the effect we are proposing to test depends on the relative size of the internal energy splitting $\mu \B$ to the particle's mass-energy $m_\I c^2$, which is larger for lighter isotopes.  On the other hand, many of the probabilities scale inversely with the trap frequency---the transition to $(n,s)=(1,-\frac12)$ scales with $\w_0^{-3}$, for example. This means that with incremental improvements in technology, the precision to which violations can be tested will increase manifold.

Additionally, one might expect that thermal noise could obscure meaningful results in a realistic setup. However, at temperatures of $\sim$890 pK, to which single Helium atoms have been cooled \cite{Manning2014}, the thermal population of the first excited state of $H$ in Eq.~\eqref{eq:FullHam} is $\sim$100 times smaller than the corresponding transition probability in Table~\ref{tab:TransitionProbs} (using the same parameters---see Appendix~\ref{sec:thermal} for details). Moreover, any further cooling would reduce this number by orders of magnitude. To a good approximation, thermal noise is limited by the smaller of the oscillator energy $\hbar \w_0$ and the Zeeman energy $\mu \B$. Reducing the former to increase the experimental signal would thus necessitate a corresponding decrease in temperature.

One ambiguous feature of our proposed experiment is the seemingly unnatural, asymmetric definition of the internal Hamiltonian in Eq.~\eqref{H-int-r}. We follow Ref.~\cite{Zych:2015vm} in choosing this definition, motivated by the constraint that the internal energy should not result in negative contributions to the systems mass. With a symmetric energy splitting, one could imagine a scenario where the external field is large enough to render the combined mass-energy negative. The choice of internal Hamiltonian has physical consequences---the symmetric definition would correspond to $a_k \to a_k - (2\lambda)^{-1}$ in our formulation, resulting in significant non-zero values for some of the transition probabilities in Table~\ref{tab:TransitionProbs}. Performing our experiment would thus quickly distinguish between the symmetric and asymmetric definitions.

Finally, the setup we have described in this Letter will surely motivate the design of more sensitive experiments to test the quantum equivalence principle. For example, a simple extension would be to consider ensembles of atoms in a single harmonic trap---the weak interactions of neutral fermions would be unlikely to affect our results significantly. Furthermore, it is well known that quantum entanglement can enhance measurement precision by a factor that scales with the square root of the number of entangled parties \cite{Giovannetti2006}; perhaps it can be employed here to overcome the low probabilities of violating events. 

{\bf Acknowledgements.---} We thank Lucas C\'eleri and Kristian Helmerson for helpful conversations. This work was supported in part by the Natural Sciences and Engineering Research Council of Canada.

\bibliography{QuantumEquivalenceRef}

\onecolumngrid
\newpage
\appendix
\section{Derivation of the Complete Hamiltonian}\label{App:H-Derivation}

The Hamiltonian for a simple harmonic oscillator in a uniform gravitational field is given by
\begin{gather}
H = \frac{P^2}{2m_\I} + \frac12 
k x^2 + m_\Gs gx,\label{eq:HAM-SHO}
\end{gather}
where we have made a clear distinction between inertial and gravitational mass. To promote this to the mass operator formalism some modifications are necessary. Typically the rest mass energy is omitted from the Hamiltonian, but its promotion to an operator warrants its inclusion, which then automatically accounts for the internal energy of the system. The Hamiltonian with mass operators is given by,
\begin{gather}
H= M_\R c^2 + \frac{P^2}{2M_\I } + \frac12\kappa x^2 + M_\Gs g x,  \label{eq:InitHam2}
\end{gather}
where the $M_k$ are the respective mass operators defined in the previous section. Inserting the definition of the mass operator, we obtain
\begin{gather}
H = \left(m_\R +\frac{1}{c^2}\hint_\R\right)c^2 + \frac{P^2}{2m_\I\left(\id + \frac{1}{m_\I c^2}\hint_\I\right) } + \frac12\kappa x^2 +  \left(m_\Gs + \frac{1}{c^2}\hint_\Gs\right)g x. 
\end{gather}
Provided that $\frac{\|\hint_\I\|}{m_\I c^2} \ll 1$, we can Taylor expand the denominator on the $P^2$ operator to first order, finding
\begin{align}
H =& m_\R c^2 +\hint_\R + \frac{P^2}{2m_\I}\left(\id - \frac{1}{m_\I c^2}\hint_\I\right)  + \frac12\kappa x^2 +  \left(m_\Gs + \frac{1}{c^2}\hint_\Gs\right)g x\nonumber\\
=& m_\R c^2 +\hint_\R + \frac{P^2}{2m_\I} + \frac 12 \kappa x^2 + m_\Gs g x + \frac{1}{m_\I c^2}\left\{-\frac{P^2}{2m_\I}\hint_\I + g x \hint_\Gs \right\}.
\end{align}
Let us now introduce the parameter $\w_0 = \sqrt{\frac{\kappa}{m_\I}}$, which is the natural frequency of the oscillator when in its lowest energy eigenstate, and complete the square for the position operator:
\begin{align}
H&= m_\R c^2 +\hint_\R + \frac{P^2}{2m_\I} + \frac 12 m_\I \w_0^2 \left(x + \nu\frac{g}{\w_0^2}\right)^2 + \frac12\frac{m_\Gs^2g^2}{\kappa} + \frac{1}{m_\I c^2}\left\{-\frac{P^2}{2m_\I}\hint_\I + g x \hint_\Gs \right\}.\label{eq:FullHam2}
\end{align}

\section{Complete first order corrected eigenstates}\label{app:FullCalc}
The unperturbed Hamiltonian $H_0$ has eigenstates $\ket{\kp{0}{n,s}}= \ket{n}\otimes\ket{s}$, where $n =0,1,2,\ldots$ are the number states corresponding to canonical momentum and position operators $P$ and $X=x+\nu g/{\w_0^2}$, and $s \in \left\{ \textstyle{+\frac12}, \textstyle{-\frac12} \right\}$ denote spin up and down respectively (we will sometimes abbreviate $ \left\{ \textstyle{+\frac12}, \textstyle{-\frac12} \right\}$ to $\{+,-\}$ in subscripts). We can express the Hamiltonian in Eq.~\eqref{eq:FullHam2} in the form
\begin{gather}
H = H_0 + \lambda V,\label{eq:Pert1}
\end{gather}
where $H_0$ is Hamiltonian with known Eigenstates, $\lambda \ll 1$ is a dimensional parameter and $V$ is the perturbation. The eigenstates of $H$ can then be expressed as a sum $\ket{\Psi_{n,s}} = \sum_{k=0}^\infty \lambda^k \ket{\kp{k}{n,s}}$, where $\ket{\kp{k}{n,s}}$ is the $k$th order correction to the eigenstate of $H_0$ with quantum numbers $n$ and $s$. We identify $\lambda = (\mu \B)/(m_\I c^2)$ to be the perturbation parameter and then define 
\begin{gather}
V = -\frac{P^2}{2m_\I} S_\I + g xS_\Gs,
\quad \mbox{where}\quad 
S_k = \frac{\hint_k}{\mu \B} = \left[\begin{array}{cc}a_k & b_k^* \\ b_k & 1 + c_k \end{array}\right]. \label{eq:SpinProjector}
\end{gather}
We need only concern ourselves with first order corrections to the eigenstates, which are given by,
\begin{gather}
\ket{\kp{1}{n,s}} = \sum_{\{l,r\}\neq \{n,s\}} \frac{\bk{\kp{0}{l,r}} {V}{\kp{0}{n,s}}} {E^{(0)}_{n,s} - E^{(0)}_{l,r}} \ket{\kp{0}{l,r}},
\label{eq:FirstOrderCorrectionFormulaApp}
\end{gather}
in which $\{l,r\}$ and $\{n,s\}$ enumerate the entire set of eigenstates of $H_0$. This can be expressed more explicitly in the forl,
\begin{gather}
\ket{\kp{1}{0,-}}= \frac{\bk{\kp{0}{0,+}}{V}{\kp{0}{0,-}}}{E^{(0)}_{0,-} -\, E^{(0)}_{0,+}}\ket{\kp{0}{0,+}}\nonumber + \displaystyle\underset{s\,\scalebox{0.7}{$\in \{-,+\}$}}{\underset{n=1}{\overset{\infty}{\scalebox{1.5}{$\displaystyle\sum$}}}}\frac{\bk{\kp{0}{n,s}}{V}{\kp{0}{0,-}}}{E^{(0)}_{0,-} - E^{(0)}_{n,s}}\ket{\kp{0}{n,s}}.
\label{eq:GroundCorr}
\end{gather}
The energy differences in the denominator are given by $E^{(0)}_{n,s} - E^{(0)}_{l,r} = \hbar\w_0(n-l) +  \mu \B\Delta_{sr}$,
with $\Delta_{+-} = 1$, $\Delta_{-+} = -1$ and $\Delta_{\pm\pm} = 0$.

Now we calculate the perturbation matrix elements. For simplicity we denote $\ket{\psi^{(0)}_{n,s}}=\ket{n,s}$ and for clarity we compute each term separately.
Expanding $P$ and $X$ in terms of creation and annihilation operators, we find
\begin{align}
\bk{k}{P^2}{n} =& \frac{m_i\hbar \w_0}{2}\left\{(2n+1)\delta_{k,n} - \sqrt{(n+1)(n+2)}\delta_{k,n+2} -\sqrt{n(n-1)}\delta_{k,n-2}\right\} \label{eq:P2Matrix}\\
\bk{k}{X}{n} =& \sqrt{\frac{\hbar}{2m_i\w_0}}\left\{\sqrt{n+1}\delta_{k,n+1} + \sqrt{n}\delta_{k,n-1}\right\}.\label{eq:XMatrix}
\end{align}
This leads to the following expression for the inertial contribution to the numerator of Eq.~\eqref{eq:GroundCorr}:
\begin{align}
\bra{k,r}{-\frac{P^2}{2m_\I }\otimes S_\I}\ket{n,s}=& -\frac{1}{2m_\I }\bk{k}{P^2}{n}\bk{r}{S_\I}{s} \nonumber \\
=& -S_{I;r,s}\frac{\hbar \w_0}{4}\left\{(2n+1)\delta_{k,n} - \sqrt{(n+1)(n+2)}\delta_{k,n+2} -\sqrt{n(n-1)}\delta_{k,n-2}\right\}. \label{eq:PertMatrixt1}
\end{align}
where $\bk{r}{S_k}{s}=S_{k:r,s}$. 

Moving to the second term, there is a slight complication. The number states belong to the operator $X=x+\frac{\nu g}{\w_0^2}$, however the perturbation appears in terms of the operator $x$. Since the two are related by a displacement the problem is easily solved:
\begin{align}
\bk{k,r}{m_\I gx\hint_\Gs}{n,s}=& S_{G;r,s}m_\I g\bra{k}{\left(X - \frac{\nu g}{\w_0^2}\right)}\ket{n}\nonumber\\
=& S_{G;r,s}m_\I g\left\{\sqrt{\frac{\hbar}{2m_\I \w_0}}\left[\sqrt{n+1}\delta_{k,n+1} + \sqrt{n}\delta_{k,n-1}\right] -\frac{\nu g}{\w_0^2}\delta_{k,n}\right\}
\label{eq:PertMatrixt2}
\end{align}

Before substituting the above expressions into Eq.~\eqref{eq:GroundCorr}, it is convenient to group terms in the correction that correspond to a change in number state, a spin flip, or both,
\begin{gather}
\ket{\psi^{(1)}_{n,s}} = \frac{\bk{\psi^{(0)}_{n,r}}{V}{\psi^{(0)}_{n,s}}}{\mu \B\Delta_{sr}} \ket{\psi^{(0)}_{n,r}} + \sum_{k\neq n} \left\{ \frac{\bk{\psi^{(0)}_{k,s}}{V}{\psi^{(0)}_{n,s}}}{\hbar\w_0(n-k)}\ket{\psi^{(0)}_{k,s}} + \frac{\bk{\psi^{(0)}_{k,r}}{V}{\psi^{(0)}_{n,s}}}{\hbar\w_0(n-k) + \mu \B\Delta_{sr}}\ket{\psi^{(0)}_{k,r}} \right\}. \label{FOwE}
\end{gather}

Beginning with the first term, we take the elements from Eqs. \eqref{eq:PertMatrixt1} and \eqref{eq:PertMatrixt2}, which correspond to $k=n$,
\begin{gather}
\bk{\kp{0}{n,r}}{V}{\kp{0}{n,s}}= -S_{I;r,s} \frac{\hbar \w_0}{4}(2n+1) - S_{G;r,s} m_\I \nu \frac{g^2}{2\w_0^2}.
\label{corr1.1}
\end{gather}
The next two terms in (\ref{FOwE}) contain a sum over all number states except $n$. Due to Kronecker Deltas, most of these terms in the sum will vanish:
\begin{align}
\sum_{k\neq n}{\frac{\bk{\kp{0}{k,s}}{V}{\kp{0}{n,s}}}{\hbar\w_0(n-k)}\ket{\kp{0}{k,s}}} =&-\frac{S_{I;s,s}}{2\hbar \w_0}\frac{\hbar \w_0}{4} \sqrt{(n+1)(n+2)} \ket{\kp{0}{n+2,s}}
+ \frac{S_{I;s,s}}{2\hbar \w_0} \frac{\hbar\w_0}{4} \sqrt{n(n-1)} \ket{\kp{0}{n-2,s}}\nonumber\\
&  - \frac{m_\I g S_{G;s,s}}{\hbar \w_0}\sqrt{\frac{\hbar}{2m_\I \w_0}}\sqrt{n+1} \ket{\kp{0}{n+1,s}}
+  \frac{m_\I gS_{G;s,s}}{\hbar \w_0}\sqrt{\frac{\hbar}{2m_\I \w_0}}\sqrt{n} \ket{\kp{0}{n-1,s}}.
\label{corr1.2}
\end{align}
The third term is almost identical to the above equation; the differences being that $\rho_{k;s,s} \mapsto \rho_{k;r,s}$ and the energy differences are larger by $\mu \B\Delta_{sr}$, as can be seen in Eq.~\eqref{FOwE}. Thus we have
\begin{align}
\label{corr1.23}
\sum_{k\neq n}&
\left[\frac{\bk{\kp{0}{k,s}}{V}{\kp{0}{n,s}}}{\hbar\w_0(n-k)}\ket{\kp{0}{k,s}}
+\frac{\bk{\kp{0}{k,r}}{V}{\kp{0}{n,s}}}{\hbar\w_0(n-k) + \mu \B\Delta_{sr}} \ket{\kp{0}{k,r}}\right]  \\\notag
&\quad\quad=-\frac{\sqrt{(n+1)(n+2)}}{4}\left(\frac{S_{I;s,s}}{2}\ket{\kp{0}{n+2,s}}+\frac{S_{I;r,s}\hbar\w_0}{\hbar\w_0 - \mu \B\Delta_{sr}}\ket{\kp{0}{n+2,r}}\right)  \\\notag
&\quad\quad + \frac{1}{4} \sqrt{n(n-1)} \left(\frac{S_{I;s,s}}{2}\ket{\kp{0}{n-2,s}}+\frac{S_{I;r,s}\hbar\w_0}{\hbar\w_0 + \mu \B\Delta_{sr}}\ket{\kp{0}{n-2,r}}\right) \\\notag
&\quad\quad- m_\I  g\sqrt{\frac{\hbar(n+1)}{2m_\I \w_0}}\left(\frac{S_{G;s,s}}{\hbar\w_0} \ket{\kp{0}{n+1,s}} + \frac{S_{G;r,s}}{\hbar\w_0 - \mu \B\Delta_{sr}} \ket{\kp{0}{n+1,r}}\right)\\\notag
&\quad\quad+ m_\I  g\sqrt{\frac{\hbar n}{2m_\I \w_0}}\left(\frac{S_{G;s,s}}{\hbar\w_0} \ket{\kp{0}{n-1,s}} + \frac{S_{G;r,s}}{\hbar\w_0 + \mu \B\Delta_{sr}} \ket{\kp{0}{n-1,r}}\right).
\end{align}

Now we can combine Eqs.~\eqref{corr1.1} and \eqref{corr1.23} to obtain an expression for the eigenstates of $H$ up to first order in $\mu \B/m_\I c^2$
\begin{align}
\ket{\Psi_{n,s}} =& \ket{\kp{0}{n,s}} + \frac{\mu \B}{m_\I c^2}\ket{\kp{1}{n,s}}\nonumber\\
=&  \ket{\kp{0}{n,s}} + \frac{\mu \B}{m_\I c^2}\Bigg\{ \frac{1}{2\mu  \B \Delta_{sr}}\Big(S_{I;r,s}\hbar \w_0(n+\frac12) + S_{G;r,s} m_\I \nu \frac{g^2}{\w_0^2}\Big)\ket{\kp{0}{n,r}} \nonumber\\
&\qquad\qquad\qquad\quad- \frac18\sqrt{(n+1)(n+2)}\left(S_{I;s,s}\ket{\kp{0}{n+2,s}}+\frac{S_{I;r,s}}{1 -\frac{\mu \B}{2\hbar\w_0}\Delta_{sr}}\ket{\kp{0}{n+2,r}}\right)  \nonumber\\
&\qquad\qquad\qquad\quad+ \frac18\sqrt{n(n-1)}\left(S_{I;s,s}\ket{\kp{0}{n-2,s}}+\frac{S_{I;r,s}}{1+ \frac{\mu \B}{2\hbar\w_0}\Delta_{sr}}\ket{\kp{0}{n-2,r}}\right) \nonumber\\
&\qquad\qquad\qquad\quad- \frac{g}{\w_0} \sqrt{\frac{m_\I(n+1)}{2\hbar \w_0}}\left(S_{G;s,s} \ket{\kp{0}{n+1,s}} + \frac{S_{G;r,s}}{1- \frac{\mu \B}{\hbar\w_0}\Delta_{sr}} \ket{\kp{0}{n+1,r}}\right)\nonumber\\
&\qquad\qquad\qquad\quad+  \frac{g}{\w_0}\sqrt{\frac{m_\I n}{2\hbar \w_0}}\left(S_{G;s,s} \ket{\kp{0}{n-1,s}} + \frac{S_{G;r,s}}{1 +\frac{\mu \B}{\hbar \w_0}\Delta_{sr}} \ket{\kp{0}{n-1,r}}\right) \Bigg\},
\label{eigenstates}
\end{align}
where $r$ denotes the opposite spin state to $s$.

\subsection{The ground state}
The ground state of this system is given by the state $\ket{\Psi_{0,-}}$ and is slightly simpler that the expression for the general state. We will need the matrix elements $S_{k;ij}$, given by
\begin{gather}
S_k = \left[\begin{array}{cc}a_k & b_k\\ b^*_k & 1 + c_k \end{array}\right],
\end{gather}
leading to
\begin{align}
\ket{\Psi_{0,-}} =&  \ket{\kp{0}{0,-}} - \frac{\mu \B}{m_\I c^2}\left[\rule{0cm}{8mm} \left(\frac{\hbar \w_0 b_\I^*}{4\mu \B} + b_\Gs^*\frac{ m_\I \nu g^2}{\mu \B\w_0^2}\right)\ket{\kp{0}{0,+}} +\frac{a_\I }{4\sqrt{2}}\ket{\kp{0}{2,-}} +\frac{b_\I^* }{4\sqrt{2}(1 + \frac{\mu \B}{2\hbar\w_0})}\ket{\kp{0}{2,+}} \right.\\ 
&\qquad\qquad\qquad \qquad\qquad\qquad \qquad + \left. \rule{0cm}{8mm}\frac{g}{\w_0}\sqrt{\frac{m_\I}{2\hbar\w_0}}\left( a_\Gs\ket{\kp{0}{1,-}} + \frac{b_\Gs^*}{1 +\frac{\mu \B}{\hbar\w_0}} \ket{\kp{0}{1,+}}\right) \right],
\label{eq:GroundStateApp}
\end{align}

We notice that if both the weak  and Einstein equivalence principles hold, then the entire correction will vanish. This is because in the ground state there is no energy which can contribute to the mass of the oscillator. This would not be that case if the oscillator were in the corrected spin-up state.

\subsection{Corrected spin-up state}\label{app:spin}
By a similar calculation we can compute the first order corrected state $\ket{\Psi_{0,+}}$, corresponding to the unperturbed oscillator being in the ground state and the spin being anti-aligned with the magnetic field. This is the first excited state when $\Delta E$ is smaller than $\hbar \w_0$. We find
\begin{align}
\ket{\Psi_{0,+}} =&  \ket{\kp{0}{0,+}} + \frac{\mu \B}{m_\I c^2}\left[\rule{0cm}{8mm} \left(\frac{\hbar \w_0 b_\I}{4\mu \B} + b_\Gs\frac{ m_\I \nu g^2}{\mu \B\w_0^2}\right)\ket{\kp{0}{0,-}} -\frac{1+c_\I }{4\sqrt{2}}\ket{\kp{0}{2,+}} -\frac{b_\I }{4\sqrt{2}(1 - \frac{\mu \B}{2\hbar\w_0})}\ket{\kp{0}{2,-}} \right.\\ 
&\qquad\qquad\qquad \qquad\qquad\qquad \qquad - \left. \rule{0cm}{8mm}\frac{g}{\w_0}\sqrt{\frac{m_\I}{2\hbar\w_0}}\left( (1+c_\Gs)
\ket{\kp{0}{1,+}} + \frac{b_\Gs}{1 -\frac{\mu \B}{\hbar\w_0}} \ket{\kp{0}{1,-}}\right) \right],
\label{eq:ExcitedState}
\end{align}
which, if we assume that all equivalence principles hold, reduces to
\begin{align}
\ket{\Psi_{0,+}}&=  \ket{\kp{0}{0,+}} - \frac{1}{mc^2}\frac{\mu \B}{4\sqrt{2}}\left(\ket{\kp{0}{2,+}} + \frac{m_\I g}{\hbar \w_0} \sqrt{\frac{\hbar}{2m_\I\w_0}} \ket{\kp{0}{1,+}}\right).
\label{EPs-hold}
\end{align}
Thus, even when all equivalence principles hold, the introduction of the mass operators introduces coupling between the oscillator states and the spin states. This is because the increase in internal energy raises the inertia of the oscillator very slightly, in accordance with Einstein's mass-energy equivalence relation. In this case, the only non-zero element of the spin part of $H^{\text{int}}$ is $\bk{\textstyle{+\frac12}}{H^{\text{int}}}{\textstyle{+\frac12}}$, so there can be no corrections to spin down states. Since only $P^2$ appears in the perturbation, the only possible terms in the perturbation series for states $\ket{\Psi_{n,+}}$ will be $\ket{\kp{0}{n+2,+}}$ and $\ket{\kp{0}{n-2,+}}$. Considering all of this and applying it to Eq.~\eqref{FOwE}, we find that the general first order corrected eigenstates are,
\begin{align}
\ket{\Psi_{n,-}} &= \ket{\kp{0}{n,-}}\\
\ket{\Psi_{n,+}} &= \ket{\kp{0}{n,+}} - \frac{\mu  \B}{mc^2} \left[ \rule{0cm}{6mm}\frac18\sqrt{(n+1)(n+2)}\ket{\kp{0}{n+2,+}} + \frac{1}{8}\sqrt{n(n-1)}\ket{\kp{0}{n-2,+}}\right.\\
& \qquad \qquad \qquad \qquad + \left.\rule{0cm}{6mm} \frac{ g}{\w_0} \sqrt{\frac{m_\I}{2\hbar \w_0}} \left(\vphantom{\frac{m_\I g}{\hbar \w_0}}\sqrt{n}\ket{\kp{0}{n-1,+}} + \sqrt{n+1}\ket{\kp{0}{n+1,+}}\right) \vphantom{\frac18\sqrt{(n+1)(n+2)}\ket{\kp{0}{n+2,+}}}\right]. \label{simplestcase}
\end{align}

\section{Other internal states} \label{app:OtherInternal}
All atomic species of the type we are proposing to use in our experiment have a rich internal structure. Since we are claiming that all internal energy couples to gravity and inertia, we must consider whether these extra levels affect our results significantly. 

If the spin degree of freedom has a set of Hamiltonians $\hint_k$ with associated perturbation Hamiltonian $V$, then we can include all other internal degrees of freedom in the Hamiltonians $\hintp_k$ with an analogous perturbation Hamiltonian $V'$. The total Hamiltonian is therefore $H= H_0 + \lambda V + \lambda' V'$, with
\begin{gather}
H_0= \hint_\R + \hintp_\R + \frac{P^2}{2m_\I } + \frac12 m_\I \w_0^2 \left( x+ \nu \frac{g} {\omega_0^2} \right)^2, \label{eq:otherH}
\end{gather}
and $\lambda = \mu \B/m_\I c^2$ as in the text and $\lambda' = \|\hintp_R\|/m_\I c^2\sim \lambda$ is a small parameter dependent on the atom's excitation energies.

We can perform the same perturbation expansion as in Appendix~\ref{app:FullCalc} and write a general perturbed state to first order as
\begin{gather}
\ket{\Psi_{n,s,r}} = \ket{\kp{0}{n,s,r}} + \lambda \ket{\kp{1}{n,s,r}} + \lambda' \ket{\kc{1}{n,s,r}} +\mathcal{O}\left(\lambda^2\right),
\end{gather}
with
\begin{align}
\ket{\kp{1}{n,s,r}} = & \sum_{\{n',s'\}\neq \{n,s\}} \frac{\bk{\kp{0}{n',s',r}} {V}{\kp{0}{n,s,r}}} {E^{(0)}_{n,s,r} - E^{(0)}_{n',s',r}} \ket{\kp{0}{n',s',r}},\nonumber \\
\ket{\kc{1}{n,s,r}} =& \sum_{\{n',r'\}\neq \{n,r\}} \frac{\bk{\kp{0}{n',s,r'}} {V'}{\kp{0}{n,s,r}}} {E^{(0)}_{n,s,r} - E^{(0)}_{n',s,r'}} \ket{\kp{0}{n',s,r'}}.
\end{align}
where $s$ and $s'$ denote the spin degree of freedom, and $r$ and $r'$ enumerate all other internal energy levels.

If we assume that $V'$ does not change appreciably when the external magnetic field is switched off, then we can write the eigenstates of $H$ in the latter's absence as 
\begin{gather}
\ket{\Psi^{\B=0}_{n,s,r}} = \ket{\kp{0}{n,s,r}} + \lambda' \ket{\kc{1}{n,s,r}} +\mathcal{O}\left(\lambda^2\right);
\end{gather}
this is the basis in which we make the final measurements in our experiment. The probability of finding the system in a given spin and oscillator state (different from the original) after turning off the field is 
\begin{align}
\sum_{r'} \left|\ip{\Psi^{\B=0}_{n',s',r'}}{\Psi_{n,s,r}}\right|^2 =& \sum_{r'} \left|\ip{\kp{0}{n',s',r'}}{\kp{0}{n,s,r}}+ \lambda \ip{\kp{0}{n',s',r'}}{\kp{1}{n,s,r}} \vphantom{\lambda' \ip{\kp{0}{n',s',r'}}{\kc{1}{n,s,r}} + \lambda' \ip{\kc{1}{n',s',r'}}{\kp{0}{n,s,r}}}\right. \nonumber \\
&\qquad+ \left.\vphantom{\ip{\kp{0}{n',s',r'}}{\kp{0}{n,s,r}}+ \lambda \ip{\kp{0}{n',s',r'}}{\kp{1}{n,s,r}}}\lambda' \ip{\kp{0}{n',s',r'}}{\kc{1}{n,s,r}} + \lambda' \ip{\kc{1}{n',s',r'}}{\kp{0}{n,s,r}}+\mathcal{O}\left(\lambda^2\right)\right|^2  \nonumber \\
\simeq& \sum_{r'} \left|\delta_{r r'} \lambda \ip{\kp{0}{n',s',r'}}{\kp{1}{n,s,r}} + \delta_{s s'}\lambda'\left(\frac{\bk{\kp{0}{n',s,r'}} {V'}{\kp{0}{n,s,r}}} {E^{(0)}_{n,s,r} - E^{(0)}_{n',s,r'}} + \frac{\bk{\kp{0}{n',s,r'}} {V'}{\kp{0}{n,s,r}}} {E^{(0)}_{n',s,r'} - E^{(0)}_{n,s,r}}\right)\right|^2
\nonumber \\
=& \lambda^2 \left|\ip{\kp{0}{n',s',r}}{\kp{1}{n,s,r}}\right|^2 ,
\end{align}
which, up to second order in $\lambda$, is exactly the form of the probabilities in Table~\ref{tab:TransitionProbs} (we have neglected higher order terms in the second line). 

\section{Thermal Noise} \label{sec:thermal}
The energy of the eigenstates of the complete Hamiltonian in eq \eqref{eq:FullHam} to first order are
\begin{gather}
E_{n,s} = E_{n,s}^{(0)} + \frac{1}{m_\I c^2} \bk{\kp{0}{n,s}}{V}{\kp{0}{n,s}},\label{eq:EnergyCorrections}
\end{gather}
where $\ket{\kp{0}{n,s}}$ are the unperturbed eigenstates of $H_0$ and $V$ is the introduced perturbation.
\begin{align}
E_{n,-} &= \bk{\kp{0}{n,-}}{H_0}{\kp{0}{n,-}} + \frac{1}{m_\I c^2}\bra{\kp{0}{n,-}}
{\left(-\frac{P^2}{2m_\I}\hint_\I + m_\I g x \hint_\Gs\right)}\ket{\kp{0}{n,-}}\\
&= \hbar \w_0(n+1/2) + \frac{\mu \B}{m_\I c^2}\left\{\frac{-m_\I \hbar \w_0
a_\I}{2m_\I} (n+1/2) + m_\I g a_\Gs \bra{\kp{0}{n,-}}{\left(X - \nu\frac{g}{\w_0^2}\right)} \ket{\kp{0}{n,-}} \right\}\\
&= \hbar \w_0(n+1/2) - \frac{\mu \B}{m_\I c^2}\left\{\frac{ \hbar \w_0 a_\I}{2} (n+1/2) + m_\I\nu\frac{g^2}{\w_0^2} a_\Gs \right\}\\
&= \hbar \w_0\left(1 - \frac{\mu \B a_\I}{2m_\I c^2}\right)(n+1/2) -  \frac{\mu  \B }{c^2}\frac{\nu g^2}{\w_0^2}a_\Gs \label{eq:CorrectedEnergyPart1}
\end{align}
Similarly for the excited spin state
\begin{gather}
E_{n,+}= \hbar \w_0\left(1 - \frac{\mu \B (1+ c_\I)}{2m_\I c^2}\right)(n+1/2) + \mu \B\left( 1 - \frac{\nu g^2}{c^2 \w_0^2}(1+ c_\I)\right). \label{eq:CorrectedEnergyPart2}
\end{gather}
The partition function is given by 
\begin{gather}
Z = \sum_{n=0}^\infty e^{-\beta E_{n,-}} + e^{-\beta E_{n,+}}\label{eq:PartFunc}
\end{gather}
We compute the quantity
\begin{align}
\sum_{n=0}^\infty e^{-\beta( \alpha + \gamma(n+\frac12))} &= e^{-\beta \alpha} \e^{-\beta \gamma/2}\sum_{n=0}^\infty \left(e^{-\beta\gamma}\right)^n\\
&= \frac{e^{-\beta \alpha} e^{-\beta \gamma/2}}{1-e^{-\beta \gamma}}
= \frac{e^{-\beta \alpha}}{2\sinh(\beta\gamma/2)},\label{eq:GeneralFormPartition}
\end{align}
where we have assumed convergence of geometric series ($|\gamma|<1$) in line two. Applying this to equation \eqref{eq:PartFunc} we obtain
\begin{gather}
Z =  \frac{\exp\left[\beta   \frac{ \mu \nu \B g^2}{c^2 \w_0^2}a_\Gs \right]}{2\sinh\left(\frac12\beta\hbar \w_0 \left(1- \frac{\mu \B a_\I}{2m_\I c^2}\right)\right)} + \frac{\exp\left[\beta \left( \frac{\mu \nu \B g^2}{  c^2\w_0^2}(1+ c_\I) - 1\right) \right]}{2\sinh\left(\frac12\beta\hbar \w_0 \left(1 - \frac{\mu \B (1+ c_\I)}{2m_\I c^2}\right)\right)}.
\end{gather}
The statistics given by the above distribution closely resemble those of the unperturbed case. It is sufficient to ignore the effects of the mass operator in order to get an estimate of thermal noise. At temperatures around 100 picoKelvin the probability of being in a state other than the ground state is less then $10^{-300}$. 

\end{document}